\documentstyle[11pt,newpasp,epsf,twoside]{article}
\def\edcomment#1{\iffalse\marginpar{\raggedright\sl#1\/}\else\relax\fi}
\reversemarginpar

\begin{document}
\title{VLA HI and OVRO CO Interferometry of a Tidal Dwarf Galaxy}
 \author{Elias Brinks}
\affil{INAOE, Apdo.\ Postal 51 \& 216, Puebla, Pue 72000, Mexico}
\author{Pierre--Alain Duc}
\affil{Service d'Astrophysique, CEA/DSM/DAPNIA, Saclay, F--91191, France}
\author{Fabian Walter}
\affil{NRAO--AOC, P.O. Box O, Socorro, NM 87801, USA}

\begin{abstract}

We present high resolution interferometric observations of the cool
atomic and cold molecular ISM of the TDG candidate Arp\ 245N, an
object resembling a dwarf galaxy in the northern tidal tail of the
interacting system NGC\,2992/3. We observed the HI line with the NRAO
VLA and the CO(1$\to$0) transition with the OVRO millimeter
interferometer at $5''-6''$ angular resolution (750\,pc linear
resolution). These datacubes offer the required spatial and velocity 
resolution to determine whether the mass concentration
near the tip of the  tail is a genuine feature, and hence a good TDG
candidate, or an artefact caused by a fortuitous alignment of our line
of sight with the direction of the  tail.
A preliminary analysis seems to confirm that  Arp\,245N
is a self--gravitating entity.

\end{abstract}
\index{o:NGC 2992}
\index{o:NGC 2993}
\index{o:Arp 245N}

\section{Introduction}

Tidal Dwarf Galaxies (TDGs) are objects resembling actively star
forming dwarf galaxies and are assembled from the debris (tidal tails
and bridges) launched into the IGM by violent galaxy interactions in
which at least one member is a gas--rich galaxy. They are composed of
stars and gas from the outskirts of one or both of the parent galaxies
involved in the interaction. 
The recent surge of interest in TDGs started with papers by Mirabel,
Lutz, \& Maza (1991) on the Superantennae and Mirabel, Dottori, \&
Lutz (1992) on the Antennae (see also the review by Duc \& Mirabel
1999). Several groups of authors have since embarked on the exciting
topic of TDGs as witnessed by these proceedings (see e.g., the
contributions by Duc et al.\ and by Braine et al.\, this volume).

Currently outstanding questions are: (i) Are TDGs really recycled
objects made of collisional debris or pre-existing galaxies involved in 
a three--body interaction? (ii) Are  TDG  genuine density enhancements in 
the tidal tails or are they merely due to projection effects along the
line of sight? 
(iii) Do TDGs form self gravitating entities or are they simple transient
 condensations? (iv) Are TDGs Dark Matter (DM) dominated, like galaxies in 
general, and dwarf galaxies in particular, or are they nearly devoid of DM, 
as theory predicts? (v) Finally do TDGs leave the potential well of their
progenitors and  
hence constitute a sizeable fraction of the known dwarf galaxy population or 
do they eventually fall back and merge, leaving no trace? 

All these questions have actually been raised for the particular TDG candidate
identified in the northern tidal tail of Arp\,245, an interacting  system 
composed of two spiral galaxies, NGC 2992 and NGC 2993.
Although Arp\,245N was observed at many wavelengths and is one of the
best studied  
TDG candidates,  its  nature as a tidal object or as a real entity have  been 
challenged.  
Smith \& Struck (2001) argued that TDG Arp\,245N could actually be 
a preexisting edge-on disk galaxy that is interacting with the other
two galaxies. 
Hibbard et al.\ (this volume) point out that this system is viewed from an
unfavorable perspective, making the projection effects particularly severe.

The combination of high--resolution  kinematical and morphological data  is 
critical to tackle all these issues. We have therefore carried out 
HI and CO interferometric observations of the system.

\section{The Interacting System Arp\,245 (= NGC\,2992/3)}
\index{o:Arp 245 (= NGC 2992/3)}

NGC\,2992/3 is a relatively nearby system, at an adopted distance of
31 Mpc (V$_{\mathrm sys} = 2311$\,km\,s$^{-1}$). Its prominent
northern tidal tail hosts a tidal dwarf galaxy candidate which because
of its proximity can be studied in detail (Fig.~1). The system was
observed by us in HI with the NRAO\footnote{The National Radio
Astronomy Observatory is a facility of the National Science Foundation
operated under cooperative agreement by Associated Universities, Inc.}
Very Large Array (VLA) at an intermediate angular resolution of $19''
\times 14''$ (Duc et al.\ 2000).

\begin{figure}[t]
\plotone{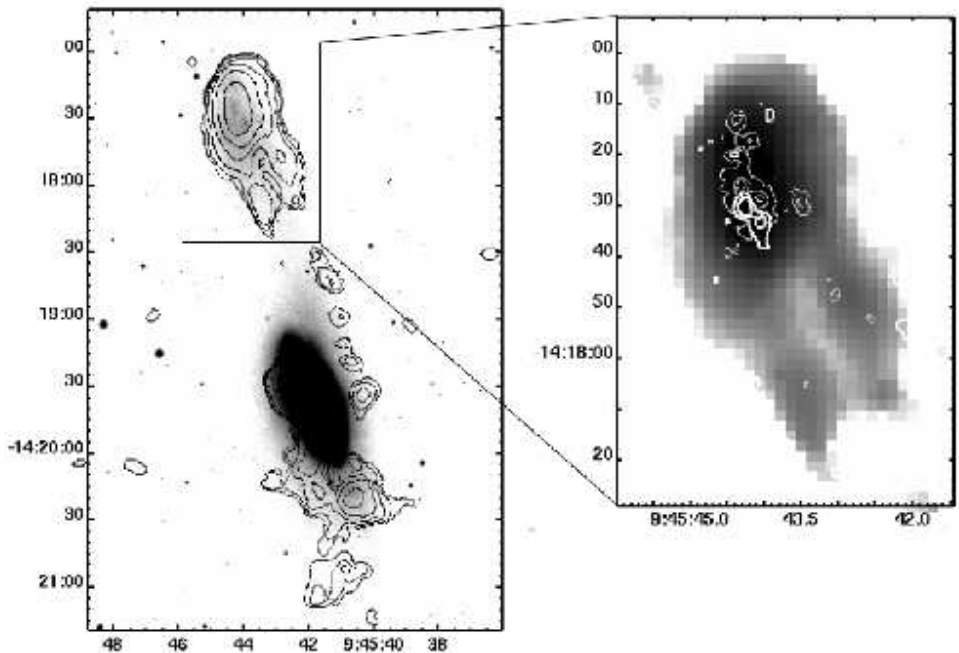}
\caption{{\em Left:\/} Optical R-band image of NGC 2992  with overlaid
HI contours from the high resolution VLA B--array observations. {\em
Right:\/} Grey 
scale representation of the HI data
centred on Arp\,245N with superimposed the H$\alpha$ emission
(thin contours) and the high resolution OVRO CO(1-0) (thick
contours). Coordinates are for a J2000.0 equinox.}
\end{figure}

Toward NGC\,2992, the HI shows a peak at the location of the tidal
dwarf candidate located at the tip of the tidal tail emanating from
NGC\,2992. HI is seen in absorption against the radio continuum from
the nucleus. NGC\,2992 is classified as a Seyfert 1.9 galaxy and in
the light of H$\alpha$ a biconical structure can be seen which extends
out into the halo. Toward NGC\,2993, the HI map has a ringlike
structure which in its western extension has no optical
counterpart. Within NGC\,2993 and Arp\,245N, HII regions are
concentrated within HI clumps and trace star--forming
regions. Numerical simulations of the NGC 2992/3 collision 
indicate that we see the system $\sim 100$\,Myr after closest approach 
(Duc et al.\ 2000).

As reported by Braine et al.\ (2000, 2001), Arp\,245 was detected with
the IRAM 30--m dish in both the CO(1$\to$0) and the CO(2$\to$1)
transitions. The observations also revealed that the CO emission is
extended along the TDG. The H$_2$ mass, assuming the standard
(Galactic) CO to H$_2$ conversion factor of N$_{\mathrm {H_2}/I_{CO}} = 2
\times 10^{20}$\,cm$^{-2}$/K\,km\,s$^{-1}$ is $\sim 1.5 \times
10^8$\,M$_\odot$ .

An oxygen abundance of 12+log(O/H)=8.6 was measured in the HII
regions of the TDG candidate. This high metallicity excludes the hypothesis
that the TDG is in fact a preexisting dwarf galaxy. It is also unlikely that
a more massive galaxy was directly involved in the interaction.
Indeed, the morphology of the system matches very well that of  the 
numerical model in which only two galaxies are interacting.

\section{Observations}

\subsection{HI Observations}

The field containing the interacting pair NGC\,2992/3 was observed
with the VLA in its B--configuration on 24 April and 7/8 May 2001 for
a total of almost 10 hours. Data calibration and reduction followed
standard procedures using the Classic AIPS data reduction package. The
final data products have a resolution of $6''$ (at 5\,km\,s$^{-1}$
velocity resolution) and reach an rms noise of 0.5 mJy\,beam$^{-1}$
per channel.

\subsection{CO Observations}

We observed the tidal dwarf near NGC\,2992 in the CO(1$\to$0)
transition using the Owen's Valley Radio Observatory's mm array (OVRO)
in the E, C and L configurations for a total of 9 tracks from October
2001 through May 2002. The equatorial E configuration was needed to
improve the beam shape for this $\delta = -14^{\circ}$ source. About
40 hours were spent on source.  Data were recorded using a correlator
setup resulting in velocity resolutions of 5\,km\,s$^{-1}$ (after
online Hanning smoothing) with a total bandwidth of 320
\,km\,s$^{-1}$. A datacube was produced using the {\sc miriad}
software package which was {\sc clean}ed to a level of about twice the
rms noise (noise: 16\,mJy\,beam$^{-1}$ in a 5\,km\,s$^{-1}$ wide
channel). The final beam size is $7.2'' \times 4.0''$.

\section{Results}

Figure 1  shows an integrated HI map of the B--array
VLA data only. It should be noted
that an interferometer acts as a spatial filter and that the B--array
is sensitive to structures with typical sizes of between $5''$ and
$120''$. As a result, applying a {\sc clean} algorithm can leave some
low level striping which will disappear once the new data are
incorporated with the intermediate resolution maps.

The right panel in Figure 1 shows as contours overlaid on the HI map,
the CO and H$\alpha$ emission. CO
is clearly detected in the OVRO observations and is found to be
resolved. The total integrated flux  is
$\sim$3\,Jy\,km\,s$^{-1}$ ($\sim10$\,K\,km\,s$^{-1}$), corresponding
to a total molecular gas mass, assuming a standard conversion factor
of N$_{\mathrm {H_2}/I_{CO}} = 2 \times
10^{20}$\,cm$^{-2}$/K\,km\,s$^{-1}$, of M$_{\mathrm
{H_2}}\sim3.5\times10^7$\,M$_{\odot}$. This is about 25\% of the
single dish flux which implies that most of the molecular gas is
distributed smoothly across the region.

\begin{figure}[t!]
\plotone{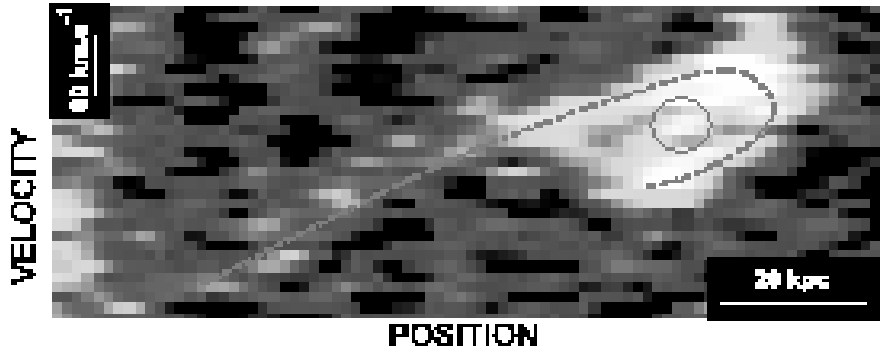}
\caption{Position--velocity diagram taken through the high
resolution HI data along a band connecting NGC\,2992 and
Arp\,245N, along the northern tidal tail.}
\end{figure}

\begin{figure}[t!]
\plotfiddle{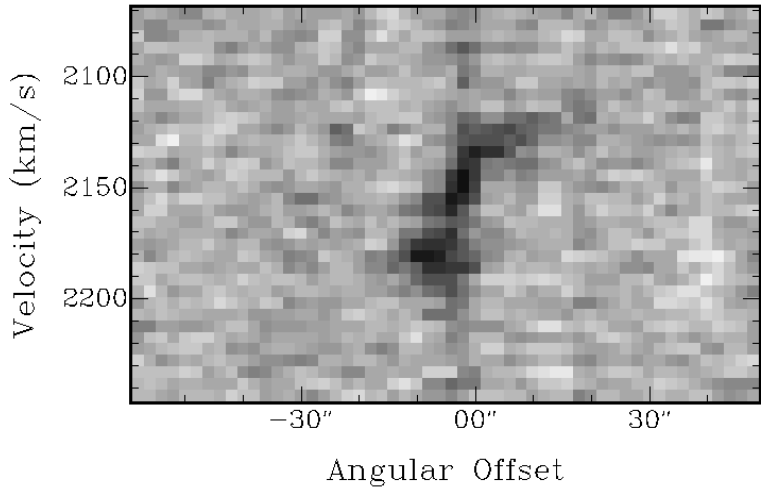}{6.5cm}{0}{40}{40}{-150}{-10}
\caption{Position--velocity diagram taken through the high resolution
HI data along a cross--cut through the TDG Arp\,245N and perpendicular
to the tidal tail. Angular distances along the cross--cut are in
arcmin. The faint vertical stripe is an artefact (see text).}
\end{figure}

\section{Discussion and Summary}
Arp\,245N is a typical TDG candidate in the sense that it is a major
HI concentration associated with recent star formation which resides
near the  tip of the tidal tail. Because the latter is seen close to
edge--on, the question thus arises whether the apparent
concentration is a genuine feature (Hibbard \& Mihos, this volume). 
Bournaud et al.\ (2003, 2004; see also the contributions by
Amram et al., this volume) have run extensive numerical models and
shown the characteristic shape in position--velocity space of a {\em
bona fide} TDG and that of a spurious feature. They show that the
kinematical signature of
projection effects is a change in the sign of the velocity gradient along
the tail before reaching its apparent tip. For curved tails that are extended 
enough in 3D space, a loop-like feature may even be seen in a
position--velocity (pV) diagram along the tails.

 Figure 2 shows such a pV diagram along the tidal tail
connecting the TDG candidate and NGC\,2992 using the intermediate
and high resolution VLA HI data. The signal was actually integrated over
a band with a width similar to that of the tail. The loop expected for
projection effects is clearly seen on the figure. The part of the tail
that is seen curving back towards NGC 2992 (as seen projected on the sky) is 
actually consistent with our earlier numerical simulations of the system
(Duc et al., 2000; see the face--on view in their Fig.~10).

The HI morphology of the tidal tail, as seen at high resolution
in Fig.\ 1, may also give some clues as to the geometry of the system.
Its U--like shape could be interpreted as being due to bending of the
tidal tail near its apparent tip. The tail is actually not seen perfectly
edge--on (as indicated, in the optical,  by the large width of the
stellar tail). 
Thanks to the higher spatial resolution provided by the VLA in its
B--configuration, we can hence 'resolve' the projection effects -- which
was impossible with the early C--array data. On this map, the 
HI column density seems to peak in the part of the tail which points
back to NGC 2992. This is where OVRO detected the bulk of the molecular
gas and where the brightest HII regions in the tail are found. The velocities
of all these phases match. The spatial and velocity coincidence between
the CO, H$\alpha$ and HI emission peaks at this location in  the tail 
is a strong indication  that a genuine condensation  is present there
and that this is likely the progenitor of a Tidal Dwarf galaxy.

At the same position, a pV diagram perpendicular to the tidal tail 
shows  a small scale velocity gradient similar to that expected
for a rotating body (see Fig.~3). The peak--to--peak
velocity range is 100\,km\,s$^{-1}$. 
A word of caution is warranted here, though. As Duc et al.\ (2000)
mentioned, the  simulated pV diagram along the same direction  in the 
numerical model  shows a similar gradient. 
Further simulations are required to disentangle the
embedded TDG from the rest of the tail. We should then be able
to determine its dynamical mass and, comparing it with the
luminous mass (corresponding to the HI condensation in the
B--array), probe its dark matter content. Not taking into account
  the line of sight crowding, and considering all the matter present at the
apparent tip of the tail, one derives a dynamical mass similar
to the luminous one and equal to $\sim 2 \times 10^9$\,M$_\odot$.
For the above-mentioned reasons, these are most likely 
overestimates.

In summary, a first analysis of new high resolution HI and CO datacubes
tends to  support the existence of a bound entity within the northern 
tail of  Arp\,245. However, they also show the kinematical signature 
expected when  part of the  tail is  bending away along the line of sight,
and eventually back to NGC 2992. 
Because of these projection effects, the size and mass of the embedded
TDG candidate derived from low resolution data are probably overestimates.

\acknowledgments{EB gratefully acknowledges a travel grant awarded by
the IAU and partial financial support in the form of CONACyT grant
No.\ 27606--E.}


\begin{references}

\reference Bournaud, F., Duc, P.--A., \& Masset, F. 2003, \aap, 411, L469

\reference Bournaud, F., Duc, P.--A., Amram, P,  \& Combes, F.  2004,
submitted to \aap   

\reference Braine, J., Duc, P.--A., Lisenfeld, U., Charmandaris, V.,
	Vallejo, O., Leon, S., Brinks, E. 2001, \aap, 378, 51 

\reference Braine, J., Lisenfeld, U., Duc, P.--A., Leon, S. 2000,
	Nature, 403, 867  


\reference Duc, P.--A., Brinks, E., Springel, V., Pichardo, B.,
	Weilbacher, P., \& Mirabel, I. F. 2000, \aj, 120, 1238 

\reference Duc, P.--A. \& Mirabel, I. F. 1999, in IAU Symp.\ 186,
	 Galaxy Interactions 	at Low and High Redshift, ed.\
	 J. E. Barnes \& D. B. Sanders 
	 (Dordrecht: Kluwer),  p. 61

\reference Mirabel, I.F., Dottori, H., \& Lutz, D. 1992, \aap, 256, L19

\reference Mirabel, I.F.,  Lutz, D., \& Maza, J. 1991, \aap, 243, 367

\reference Smith, B.J., \& Struck, C. 2001, \aj, 121, 710

\end{references}
\end{document}